\documentclass[showpacs,aps,prd,twocolumn,floats,superscriptaddress]{revtex4}

\usepackage{graphicx}
\usepackage{amsfonts}
\usepackage{amsmath}
\usepackage{subfigure}
\usepackage{bm}

\def \st{\tilde{s}}
\def \sh{\hat{s}}

\def \A {{\cal A}}

\def \N {{\cal N}}
\def \Nind {N_{\rm ind}}

\def \e0{\epsilon_0}

\def \h {{1 \over 2}}

\def\be{\begin{equation}}
\def\ee{\end{equation}}
\def\bea{\begin{eqnarray}}
\def\eea{\end{eqnarray}}
\def \no {\nonumber}
\def\lsim{\mathrel{\rlap{\lower4pt\hbox{\hskip1pt$\sim$}}
    \raise1pt\hbox{$<$}}}                
\def\gsim{\mathrel{\rlap{\lower4pt\hbox{\hskip1pt$\sim$}}
    \raise1pt\hbox{$>$}}}                
\def \old {{\rm old}}
\def \new {{\rm new}}
\def \p {{(+)}}
\def \Lo {{\Lambda_\old}}
\def \Los {{\Lambda^*_\old}}
\def \Ln {{\Lambda_\new}}
\def \Lns {{\Lambda^*_\new}}

\begin{document}

\preprint{OUTAP 275, IUCAA 05/2007}

\title{Detecting gravitational waves from inspiraling binaries with a network of detectors :
coherent strategies by correlated detectors}

\author{Hideyuki Tagoshi}
\affiliation{Department of Earth and Space Science,
Graduate School of Science, Osaka University, Toyonaka,
Osaka 560-0043, Japan}

\author{Himan Mukhopadhyay}
\affiliation{Inter-University Centre for Astronomy and Astrophysics,\\
Post Bag 4, Ganeshkhind, Pune 411007, India}

\author{Sanjeev Dhurandhar}
\affiliation{Inter-University Centre for Astronomy and Astrophysics,\\
Post Bag 4, Ganeshkhind, Pune 411007, India}

\author{Norichika Sago}
\affiliation{School of Mathematics, University of Southampton,
Southampton SO17 1BJ, United Kingdom}

\author{Hirotaka Takahashi}
\affiliation{Max-Planck-Institut f\"{u}r Gravitationsphysik,
Albert-Einstein-Institut, Am M\"{u}hlenberg 1, D-14476 Golm bei Potsdam,
Germany}

\author{Nobuyuki Kanda}
\affiliation{Department of Physics, Graduate School of Science, Osaka City University,
Osaka 558-8585, Japan}


\begin{abstract}
We discuss the coherent search strategy to detect gravitational waves from inspiraling compact
binaries by a network of correlated laser interferometric detectors.
From  the maximum likelihood ratio statistic, we obtain a coherent statistic 
which is slightly different from and generally better than what we obtained in our previous work.
In the special case when the cross spectrum of two detectors
normalized by the power spectrum
density is constant, the new statistic agrees with the old one.
The quantitative difference of the detection probability for a given false alarm rate
is also evaluated in a simple case.
\end{abstract}

\maketitle

\section{Introduction}
\label{intro}

Gravitational waves (GW) is one of the most important predictions of general relativity.
In order to detect GW, several GW detectors are currently in operation around the world.
However, the direct detection of GW has not been possible so far.
Since the expected amplitude of GW is very small, multi-detector searches for GW are 
very important. A multi-detector search would (i) improve the detection efficiency;
(ii) improve our confidence in detection of a GW event, 
if a candidate event is registered; (iii) provide useful directional information on  the GW event if the detectors are sufficiently geographically separated; (iv) provide polarization information if differently oriented.

In our previous paper \cite{MSTDTK} (hereafter, Paper I), we discussed two different multi-detector detection strategies; the coherent strategy and the coincident strategy.
The coherent strategy was discussed by Pai, Dhurandhar and Bose \cite{PDB}, and by Finn \cite{Finn}. The target signal was the GW from inspiraling compact binaries for which the wave forms have been well studied. We considered the situation of two co-aligned and co-located detectors.
We constructed the coherent statistic based on the maximum likelihood method.
We then compared the coherent method with the conventional coincident method by comparing
the receiver operating characteristic (ROC) curves. We found that we could obtain better detection probability by the coherent method than by the coincident method.
This conclusion was not changed even if the two detectors were correlated.

In this paper, we consider two detectors with the same
configuration as in Paper I. We discuss the refined derivation of
the coherent statistic from the maximum likelihood statistic in
the case when the detectors' noises are correlated. This statistic
is slightly different from, and better than the one, we
obtained in our previous work (Paper I). Here we present the mathematical proof of the
superiority of the new statistic over the old one. In
the special case when the cross-spectrum of two detectors
normalized by the power spectrum density is constant, the new
statistic agrees with the old one. Further, the quantitative difference of
the detection probability for a given false alarm rate is also
evaluated for a simple case.

\section{The new coherent statistic}
\label{coherent}

We consider two detectors which have same orientation, same arm length
and the same noise power spectral density.
We denote the one sided noise power spectrum density (PSD) common to each of the detectors
as $S_h(f)$ which has the property,
\be
\langle n_I (f) n^{*}_I(f') \rangle = \h  S_h (f) \delta (f - f') \,,
\label{psd}
\ee
where $n_I(f)$ is the noise in detector $I = 1,2$. The angular brackets denote ensemble average.
\par
We assume the correlation of two detectors expressed by
\be
\langle n_1 (f) n^{*}_2(f') \rangle = \h \epsilon (f)  S_h (f) \delta (f - f') \, ,
\label{noisecorr}
\ee
where $\epsilon (f)  S_h (f)$ represents the cross-spectrum. The function, $\epsilon(f)$, is the cross-spectrum normalized by the power spectral density. We assume, for simplicity, the cross-correlation in the time domain is a function of $|t-t'|$. Thus, $\epsilon (f)$ is real.

The response of a detector $I$ to a gravitational wave from an inspiraling compact binary
in the frequency domain is written as:
\bea
\st_I(f)={\mathcal{N}} E_I (\phi, \theta, \psi; \iota) f^{-7/6} 
e^{i\Psi(f; t_c, \delta_c, \tau_0, \tau_3)} ,
\label{eq:stilde}
\eea
where $t_c$ and $\delta_c$ are respectively the coalescence time and the coalescence phase of the binary,
$\tau_0$ and $\tau_3$ are defined through mass parameters by Eq. (2.6) in Paper I.
$E_I(\phi, \theta, \psi; \iota)$ is the extended antenna pattern function
which depends on the orientation angles. The definitions of $E_I$ and the amplitude normalization  $\mathcal{N}$ are given in Paper I.

The function $\Psi(f; t_c, \delta_c, \tau_0, \tau_3)$ describes the phase evolution
of the inspiral waveform. We adopt the 3PN formula given by
\begin{eqnarray}
\nonumber
&& \! \! \! \! \! \! \! \! \! \Psi(f;t_c,\delta_c,\tau_0,\tau_3)  = 2 \pi f t_c - \delta_c - \frac{\pi}{4}\\
&& \! \! \! \! \! \! \! \! \! \! \! \!+ \frac {3}{128 \eta} (\pi M f)^{-5/3} \sum_{k=0}^{6} \alpha_k (\pi M f)^{k/3} \, ,
\label{phase}
\end{eqnarray}
where $M$ is the total mass, $\eta$ is the ratio of the reduced mass to the total mass, 
and $\alpha_k$ are given by Eq. (2.8) in Paper I.

The form of (\ref{eq:stilde}) allows us to write the explicit quadrature representation
of the $i$-th template as,
\bea
\no
\tilde{s}_I(f;\vec \mu_i, t_c, \delta_c) &=& \A (\st_0 (f; \vec \mu_i, t_c) \cos \delta_c \\
                                    && + \st_{\pi/2}(f; \vec \mu_i, t_c) \sin \delta_c) \, ,
\label{eq:quadrature}
\eea
where
\bea
&&\st_0 (f; \vec \mu_i, t_c)= 
a^{-1} f^{-7/6} e^{i\Psi(f; t_c, \delta_c=0, \tau_0, \tau_3)} , \no\\
&&\st_{\pi/2}(f; \vec \mu_i, t_c)=-i \st_0 (f; \vec \mu_i, t_c), \no\\
&&a=\left[4 \int_{f_l}^{f_u} df \frac{f^{-7/3}}{S_h(f)} \right]^{1/2}, \\
&&\A=\N E a \,. \label{eq:A}
\eea
Because the detectors are identically oriented, $E_I = E$. The templates $s_0$ and $s_{\pi/2}$ have been normalized such that $(s_0,s_0) = (s_{\pi/2}, s_{\pi/2})  = 1$, where the scalar product $(a,b)$ of two real functions $a(t)$ and $b(t)$ is defined as:
\be
\left ( a, b \right ) = 2 \int_{f_l}^{f_u} df \ \frac{  \tilde{a}(f) \tilde{b}^*(f) \ + \
\tilde{a}^*(f) \tilde{b}(f)}{S_h(f)} \,.
\label{eq:scalar}
\ee
The quantities $f_l$ and $f_u$ are respectively, the lower and upper cut off frequency.

The output data from each detector, $x_I$ ($I=1,2$), can be written as
$x_I=n_I+s_I$, 
where $n_I$ is noise and $s_I$ is gravitational wave signal.
We define two pseudo data by
\bea
x_{\pm}&=&(x_1\pm x_2)/\sqrt{2} \no \\
&=& n_\pm+ s_\pm,\\
n_\pm&\equiv& (n_1\pm n_2)/\sqrt{2}, \\
s_\pm&\equiv& (s_1\pm s_2)/\sqrt{2}.
\eea
The two noise, $n_\pm$, have property
\bea
\langle n_\pm (f) n^{*}_\pm(f') \rangle&=&\frac{1}{2}(1 \pm \epsilon(f))  S_h (f) \delta (f - f') \no\\
&\equiv& \frac{1}{2} S_{h(\pm)}(f) \delta (f-f') \, ,\\
\langle n_+ (f) n^*_- (f)\rangle&=&\langle n_+^* (f) n_- (f)\rangle=0 \,. \label{n+n-}
\eea
Since $n_I(f)$ are Gaussian variables, $n_\pm(f)$ are also Gaussian.
Then, from Eq. (\ref{n+n-}), we find that $n_+(f)$ and $n_-(f)$ are independent each other.
Thus, it is straightforward to find the probability distribution function of $n_\pm(t)$.
We note that the signal, $s_-$ is identically zero since the two detectors have the same orientation, the same arm length. We thus need to consider $x_+$ only.

We derive the likelihood ratio from $x_+$. We denote the probability distribution function of
$x_+(t)$ in the absence of signal as $P(x_+|0)$. The probability distribution function of 
$x_+(t)$ in the presence of the signal $s_+(t)$ is given by $P(x_+|s_+)=P(x_+-s_+|0)$.
Since $x_+$ is Gaussian in the absence of signal, the logarithm of the likelihood ratio,
$\lambda_+$ is given by,
\bea
\ln\lambda_+&=&\ln \frac{P(x_+|s_+)}{P(x_+|0)} \no\\
&=&(x_+,s_+)_{(+)}-\frac{1}{2}(s_+,s_+)_{(+)},
\eea
where the new inner product, $(a,b)_{(+)}$, for real functions,
$a(t)$ and $b(t)$ is given by,
\bea
(a,b)_{(+)}=2\int_{f_l}^{f_u} df
\frac{\tilde{a}(f) \tilde{b}^*(f)+\tilde{a}^*(f) \tilde{b}(f)}{S_{h(+)}(f)}.
\eea

The plus signal $s_+$ is given by,
\bea
&&s_+(f)= \no\\
&&\sqrt{2}\A_\p(\sh^\p_0(f)\cos\delta_c+\sh^\p_{\pi/2}(f)\sin\delta_c),
\eea
where
\bea
&&\A_\p=\N E a_\p,\\
&&a_\p=\left[4 \int_{f_l}^{f_u} df \frac{f^{-7/3}}{S_{h(+)}(f)} \right]^{1/2},
\eea
and $\sh^\p_0$ and $\sh^\p_{\pi/2}$ are orthonormalized templates:
\bea
&&(\sh^\p_0,\sh^\p_0)_{\p}=(\sh^\p_{\pi/2},\sh^\p_{\pi/2})_\p=1,\\
&&(\sh^\p_0,\sh^\p_{\pi/2})_\p=0.
\eea

We define the normalization factor of $s_+$ as
\be
b_\p^2=(s_+,s_+)_{(+)}=2\A_\p^2.
\ee
The normalized $s_+$ is given by,
\bea
\sh_+(f)&=&\frac{1}{b_\p}s_+(f) \no\\
&=&\sh^\p_0(f)\cos\delta_c+\sh^\p_{\pi/2}(f)\sin\delta_c. 
\eea
Then the log-likelihood ratio becomes:
\be
\ln \lambda_+=b_\p(x_+,\sh_+)_\p-\frac{1}{2}b_\p^2
\ee
The maximization of $\ln\lambda_+$ with respect to $b_\p$ and $\delta_c$  (see \cite{PDB}) gives:
\bea
\max_{b_\p, \delta_c} \ln \lambda_+&=&
\frac{1}{2}[(x_+,\sh^\p_0)_\p^2+(x_+,\sh^\p_{\pi/2})_\p^2], \no \\
&\equiv& \frac{1}{2} \Lambda_\new.
\eea
We define
\bea
c_0^\p=(x_+,\sh^\p_0)_\p, \quad 
c_{\pi/2}^\p=(x_+,\sh^\p_{\pi/2})_\p,
\eea
and we obtain
\bea
\Lambda_\new&=&(c_0^\p)^2+(c_{\pi/2}^\p)^2.
\eea
The mean and variance of $c_i^\p$ ($i=0, \pi/2$) are given by
\bea
&&\langle c_0^\p \rangle=\sqrt{2}\A_\p\cos\delta_c, \\
&&\langle c_{\pi/2}^\p \rangle=\sqrt{2}\A_\p\sin\delta_c, \\
&&\langle (c_i^\p-\langle c_i^\p\rangle)^2 \rangle=1.
\eea
Since $c_i^\p$ obeys the Gaussian distribution,
the PDF of  $\Ln$ in the absence of signal ($\A_\p=0$)
is given by
$p_0(\Ln)=\exp\left(-\Ln/2\right)/2$.
This gives the false alarm probability by one template as 
\bea
P_{FA}^{\rm 1 template}&=&\int_{\Lns}^\infty d\Ln p_0(\Ln)\no\\
&=&\exp\left(-\Lns/2\right).
\eea
This is the false alarm rate in the case when the parameters of the signal
are known. However, in general, we do not know, a priori, the parameters of the signal
such as $M, \eta$, and $t_c$. We thus need to compute $\lambda_+$ for 
a bank of templates. This increases the false alarm rate.
As in Paper I, we introduce the effective number of statistically independent
templates, $\Nind$. The false alarm probability for the template bank
thus takes the form,
\bea
P_{FA}^\new=\Nind \exp(-\Lns/2). \label{pfanew}
\eea
The PDF of $\Ln$ in the presence of signal is given by
\bea
p_1(\Ln)&=&\frac{1}{2}\exp\left[-\frac{1}{2}(\Ln+2\A_\p^2) \right] \no\\
&&\quad \times I_0\left[\sqrt{2} \A_\p \sqrt{\Ln}\right].
\eea
The detection probability becomes
\bea
&&P_{DE}^\new(\Lns)=\int_{\Lns}^\infty d\Ln p_1(\Ln) \no\\
&&=\int_{\Lns}^\infty d\Ln
\frac{1}{2}\exp\left[-\frac{1}{2}(\Ln+2\A_\p^2) \right]\no\\
&&\quad \times I_0\left[\sqrt{2} \A_\p \sqrt{\Ln}\right].
\label{pdenew}
\eea

\section{Summary of Paper I}
\label{SPI}

In Paper I, we did not consider "$\pm$" data but used "$\pm$" correlations instead.
The results of the earlier analysis are the following.

The false alarm probability when the data is passed through $\Nind$ independent templates is given by
Eq.(3.15) of Paper I:
\bea
P_{FA}^{\old} (\Lambda^*_\old) = \Nind \exp \left( - \frac{\Lambda_\old^*}{2(1+\e0)} \right),
\label{pfapaperI}
\eea
and the detection probability as given by Eq. (3.16) of Paper I is:
\begin{eqnarray}
\nonumber
P_{DE}^{\old} (\Lambda^*_\old) & = & \int_{\Lambda_\old^*}^{\infty} \frac{d\Lambda_\old}{2(1+\e0)}
\hbox{exp} \left[-\frac{(\Lambda_\old + 2 \A^2)}{2(1+\e0)} \right]\\
& \times & I_0 \left (\A \frac{\sqrt{2\Lambda_\old}}{(1+\e0)} \right),
\label{pdepaperI}
\end{eqnarray}
where
\be
\e0 = \frac{4}{a^2} \int_{f_l}^{f_u} df \frac {\epsilon(f) f^{-7/3}}{S_h (f)} \,.
\ee

\section{A mathematical proof}

Now, we prove that for a given false alarm rate, the detection probability
by $\Ln$ is better than $\Lo$ in general.
First, we note that by setting $t\equiv \Lo/(1+\epsilon_0)$,
$t^*\equiv \Los/(1+\epsilon_0)$, and
$\A_0=\A/\sqrt{1+\epsilon_0}$, 
Eq.(\ref{pfapaperI}) and (\ref{pdepaperI}) are rewritten as
\bea
\nonumber
P_{FA}^\old (t^*)&=&\Nind \exp \left( - \frac{t^*}{2} \right),
\label{pfapaperIa}\\
P_{DE}^\old (t^*) & = & \int_{t^*}^{\infty} \frac{dt}{2}
\hbox{exp} \left[-\frac{1}{2}\left(t + 2 \A_0^2\right)\right] \no \\
&& \times  I_0 \left (\A_0\sqrt{2t} \right).
\label{pdepaperIa}
\eea
We find that the formulae for the false alarm rate, (\ref{pfanew}) and
(\ref{pfapaperIa}) are same. Thus, for a given false alarm rate,
we have the same threshold.
The functional forms of the detection probability,
(\ref{pdenew}) and (\ref{pdepaperIa}) are the same.
The difference comes from
the amplitude of signal, $\A_\p$ and $\A_0$
in (\ref{pdenew}) and (\ref{pdepaperIa}).

If $\mu(f)$ is a non-negative real function (a PSD for instance) we define `average' of the function $\mu$ to be:
\be
\langle \mu \rangle = \frac{4}{a^2} \int_{f_l}^{f_u} df \frac {\mu(f) f^{-7/3} }{S_h (f)} \,
\label{defn}
\ee
Then in terms of the average we may write, $\e0 = \langle \epsilon(f) \rangle$.

From the definition (\ref{defn}), it follows that \be
\frac{a_{(+)}^2}{a^2} = \langle (1+\epsilon(f))^{-1} \rangle \ee

From the Schwarz inequality, it immediately follows that,
\bea
\langle (1+\epsilon(f))^{-1} \rangle \langle (1+\epsilon(f)) \rangle &\geq & 1 \,, \no \\
\frac{a_{(+)}^2}{a^2} (1+\e0) & \geq & 1 \,. \label{schwrz} \eea

The Schwarz inequality Eq. (\ref{schwrz}) then implies $\A_\p^2
\geq \A_0^2$; the equality holding if and only if, $\epsilon (f)$
is constant. This shows that the detection probability by $\Ln$ is larger than that by $\Lo$.

\begin{figure}[htbp]
\includegraphics[width=8cm]{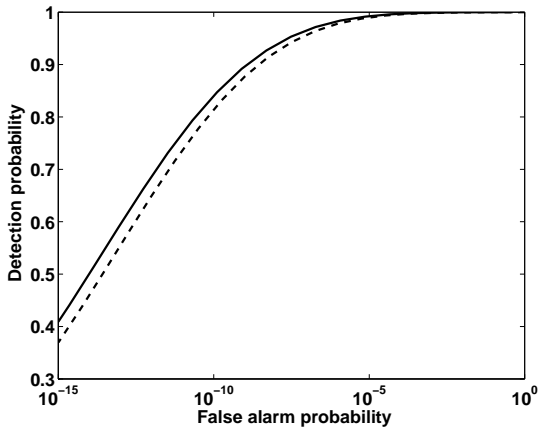}
\caption{\label{fig:1}
The ROC curve (detection probability versus false alarm probability) of the old (dashed line) and new (solid line) statistic in the case when $\epsilon(f)=0.7$ in the frequency 50Hz $\leq f \leq$ 70Hz and zero otherwise.}
\end{figure}

\section{An example}

We examine quantitatively the increase in the detection probability produced by the new statistic.
We use LIGO I noise PSD given in Paper I. We take $\Nind=6\times 10^6$ which
is approximately the value in Paper I. We set $\A=7$.
We assume the following noise correlation model:
\be
\epsilon(f)= \left\{\begin{array}{ll}
0.7 & {\rm (50Hz}\leq f \leq {\rm 70Hz)} \\
0 & {\rm otherwise}
\end{array}
\right.
\ee
In this case, we have $\epsilon_0=0.05665$,
$A/\sqrt{1+\epsilon_0}=6.809$, and $\A_\p=6.882$.
The ROC curves are shown in Fig.\ref{fig:1}. We see that the new statistic gives improved 
detection probability. For the false alarm rate of $10^{-10}$,
$P_{DE}^\old=0.814$ and $P_{DE}^\new$=0.840. The difference is 0.026.

\section{Conclusion}

We have discussed the coherent search strategy to detect gravitational waves from inspiraling compact binaries with a network of laser interferometric detectors. Based on the maximum likelihood method, we have derived a coherent statistic for detection of gravitational waves from inspiraling compact binaries by two aligned, correlated detectors.
This statistic is slightly different from the one which we obtained in our previous work.
We found that, in general, we can obtain a better detection probability for a given false alarm rate by the new statistic than by the old one. In the special case when the cross-spectrum normalized by the power spectrum density is constant, the new statistic agrees with the old one.
The quantitative difference of the detection probability for a given false alarm rate
was evaluated in a simple case, and we found that there was a significant difference.
In Paper I, we had concluded that the coherent detection strategy is better than the coincident strategy. Since we can have an increased detection probability by the new statistic,
this conclusion does not change even in the correlated case; on the contrary, the result is strengthened.

\begin{acknowledgments}
S. Dhurandhar acknowledges the DST and JSPS Indo-Japan international cooperative programme
for scientists and engineers for supporting visits to Osaka City University,
Japan and Osaka University, Japan.
H. Tagoshi and N. Kanda thank JSPS and DST under the same Indo-Japan programme
for their visit to IUCAA, Pune.
H. Mukhopadhyay thanks CSIR for providing research scholarship.
This work was supported in part by Monbu Kagakusho Grant-in-aid
for Scientific Research of Japan (Nos. 14047214, 16540251).
\end{acknowledgments}

\end{document}